\documentstyle[natbib209,aasj,mnfixes,epsfig]{mn-nat}
\begin{document}

\title[QPOs during magnetar flares]
{QPOs during
magnetar flares are not driven by mechanical normal modes
of the crust.}
\author[Levin ]{Yuri Levin 
\\
$^1$Leiden Observatory, P.O. Box 9513, NL-2300 RA Leiden, The Netherlands
\\
$^2$Lorentz Institute, P.O. Box 9506, NL-2300 RA Leiden, The Netherlands}

\date{printed \today}
\maketitle
\begin{abstract}
Quasi-Periodic Oscillations (QPOs) have been observed during 
three powerful magnetar flares, from SGR0526-66, 
SGR1806-20 and SGR1900+14. These QPOs
have been commonly interpreted as being driven by the 
mechanical modes of the magnetar's solid
crust which are excited during the flare. Here we show that
this interpretation is in sharp contradiction with the conventional
magnetar model. Firstly, we show that a magnetar crustal mode decays on
the timescale of at most a second due to the emission of Alfven waves
into the neutron-star interior. A possible modification is then to assume that
the QPOs are associated with  the magnetars' {\it global}  modes.
However, we argue that at the frequencies of the observed
QPOs, the neutron-star core is likely to support a continuum
of Magneto-Hydrodynamic (MHD) normal modes. We demonstrate this on a completely solvable
toy model which captures the essential physics of the system.
We then show that the frequency of the global mode of the whole 
star is likely to have a significant imaginary component, and its amplitude is likely
to
decay on a short timescale. This is not observed. Thus we conclude that
either (i) the origin of the QPO is in the magnetar's magnetosphere, or
(ii) the magnetic field has a special configuration: either
it is expelled from the magentar's core prior
to the flares, or it's poloidal component has very small coherence length.




\end{abstract}

\section{Introduction}

In a prophetic paper, Duncan and Thompson (1992) argued that magnetars---a class
of neutron stars with  super-strong ($10^{14}$---$10^{15}$G) magnetic 
fields---must exist. Treated at first with some skepticism,
the magnetar paradigm has proved extremely successful in
explaining the rich phenomenology of Anomalous X-ray Pulsars (AXPs) and
Soft Gamma-ray Repeaters (SGRs) (e.g., Thompson and Duncan 1995, TD). 
In particular one phenomenon which has an appealing explanation within the 
magnetar paradigm are Giant Flares from SGRs.
They are thought to be powered by an impulsive 
release of  magnetic energy stored in the neutron star. The
release may be triggered 
by a fracture in the magnetically-stressed crust (TD) or by a sudden 
reconnection in a twisted magnetosphere (Lyutikov 2003). So far 3 Giant
Flares have been observed: from SGR0526-66 (Mazets et al.~1979),
SGR1900+14 (Hurley et al~1999, Feroci et al.~1999), and SGR1806-20 (Hurley
et al.~2005, Palmer et al.~2005). In each of the flares' light-curves,
there is intriguing evidence for  QPOs with frequencies of
tens of Hertz. Firstly, the $43.5$Hz QPO was reported by
Barat et al.~(1983) in the 1979 flare from SGR0526-66. Then recently 
Israel et al.~(2005) have found several QPOs in the 2004 Giant Flare from 
SGR1806-20, with frequencies from $18$ to $90$ Hz. These QPOs 
were detected with very high signal-to-noise and lasted for 
$\sim 100$ seconds; their presence was recently
confirmed by Watts and Strohmayer (2005, WT) who used  data
from a different satellite. WT have also detected a relatively 
high-frequency QPO, of $625$Hz. Lastly, prompted by the Israel et al.~observations,
Strohmayer and Watts (2005) have found multiple QPO in the light-curve of the 
1998 Giant Flare from SGR1900+14.

The QPOs have been widely associated with  torsional modes of the neutron-star
crust [Barat et al.~(1983) made the first suggestion, followed by a 
more detailed analysis of Duncan (1998), Israel et al.~(2005),
Strohmayer and Watts (2005) and Piro (2005)]. Indeed, it is attractive
to associate a stable QPO with a mechanical mode of a neutron star,
and the frequencies of crustal torsional modes are of the right order of 
magnitude\footnote{The $625$Hz QPO found by WT is already somewhat
problematic for the crustal-mode picture. One can associate it with
a crustal shear mode which has 1 radial node (n=1; see e.g.~Piro 2005). However, a large
multitude of $l<5$, $n=1$ shear modes exist around that frequency, with
spacings of several Hz.
There is no physical reason why just one of them should be excited,
and not many.} However,  we  argue below that this interpretation is not viable 
within the magnetar paradigm. In the next section, 
we show that the crustal mode of a magnetar is not stable but decays
within seconds
by emitting Alfven waves into 
the core. This in direct contradiction with the observations.
Chris Thompson (private communications) has suggested 
that one may associate the QPOs with the  MHD-elastic modes
of the {\it whole} neutron star. However, in section 3 we argue that 
the frequency of such global mode is
likely to have an imaginary component, and  the mode
decays on a short timescale.
This is due to the continuous spectrum of MHD modes supported
by the liquid core.

\section{Decay of magnetar crustal modes.}
Magnetic fields mechanically  connect the elastic crust with the liquid core.
Both media can be considered as perfect conductors,
and ideal MHD is applicable with magnetic field lines frozen into
the media. The Alfven wave speed in the core is given by
\begin{equation}
v_a=\sqrt{BB_{\rm cr}\over 4\pi \rho \eta}\simeq 3\times 10^7
\sqrt{B_{15}\over \rho_{14} \eta} \hbox{cm/s}.
\label{va}
\end{equation}
Here $B=10^{15}B_{15}$G is the magnetic field inside the core,
$\rho=10^{14}\rho_{14}\hbox{g}/\hbox{cm}^3$ is the density,
$B_{\rm crit}\simeq 10^{15}$G is the field inside superconducting
fluxtubes, and $\eta$ is the fraction of the neutron-star core involved in the
Alfven-wave motion. If the neutrons are superfluid, then they decouple from the
Alfven wave and $\eta\sim 0.1$. If protons do not form a superconducting fluid,
$B_{\rm crit}$ should be substituted by $B$ in the above equation.

The timescale for an Alfven wave to cross the neutron-star core is
$\sim 0.05$s. Therefore, even if the oscillation was initially localized in the
crust, the rest of the core will get involved in less than a second.
It is instructive to estimate the power $W$
radiated in the Alfven waves from the
crust:

\begin{equation}
W\sim \rho\omega^2\zeta^2 v_a A,
\label{W}
\end{equation}
where $\omega$ is the angular frequency of the oscillation, $\zeta$ is the
amplitude of the crustal oscillation at the crust-core interface,
and $A$ is the inner surface area of the crust. The timescale
on which the energy is drained from the crustal mode is given by
\begin{equation}
\tau\sim{M\over \rho v_a A}\sim 0.01\hbox{s}.
\label{tau}
\end{equation}
Here $M$ is the mass of the crust. 

In a completely rigorous treatment of the problem various corrections of order 1 would
occur. Partial reflection of the Alfven waves is possible, but it will not
be a major correction since the wavelength and the relevant scaleheights
(density, magnetic pressure) are 
comparable to the neutron-star radius, and thus there is no impedance mismatch
which is needed for significant reflection. We note that the situation here is 
qualitatively different from that considered by Blaes et al.~(1989), who
considered small-wavelength shear waves propagating from the depth of the crust
into the neutron-star magnetosphere, and found large reflection coefficients.
Here we deal with global modes of the crust, which feature almost radius-independent
horizontal displacements. The motion of the inner surface of the crust, together with
frozen-in magnetic field-lines, should be considered as boundary condition
for launching the Alfven waves into the core.  

The geometry of the 
magnetic field is another obvious
factor, but for a generic configuration we do not expect a qualitative change in radiated Alfven-wave power.
If however magnetic field was confined to the crust or had a very small coherence length in its
poloidal component, then the radiated Alfven-wave power could decrease significantly. We do not 
know any compelling argument for either.
Thus we conclude that even if the crustal mode was originally  excited,
magnetic stresses would significantly reduce the mode amplitude in a 
fraction of a second and redistribute the energy within a liquid core.
Therefore, barring special magnetic-field geometry, 
a pure crustal mode cannot be associated with a stable QPO 
during the flare. One may hope to associate the {\it global}
MHD-elastic mode with the QPO. This 
proposal (due to Chris Thompson) runs
into potential difficulties which we discuss
in the next section.

\section{Global modes.}
Computing global MHD modes of a  star is clearly a difficult task.
In recent tour-de-force study  Reese, Rincon, and Rieutord (2004, RRR) have
found the eigenmodes  of a shell of a conductive incompressible fluid with
a dipole magnetic field.  
[see also Rincon and Rieutord 2003, and Levin and D'Angelo (2004)]. 
However, we are not just limited by  purely computational difficulties.
In a magnetar core
the magnetic pressure is sub-dominant, and there is virtually no conversion of
the Alfven waves into magnetosonic waves. The core responds to the wave as an
incompressible fluid, and the wave propagates along the field lines. Since different
field lines have different lengths in the core, and different average 
Alfven velocities along them, one might suspect 
that the core supports a continuum of the
Alfven modes. This is hard to prove generally, but below we, in the
spirit of Appendix B in RRR, present a toy-model
example where the Alfven-mode spectrum can be computed explicitly.

Consider a perfectly conducting incompressible fluid sandwiched in a box,
with top and bottom plates also being  perfect conductors; see Figure 1.
\begin{figure}
\begin{center}
\epsfig{file=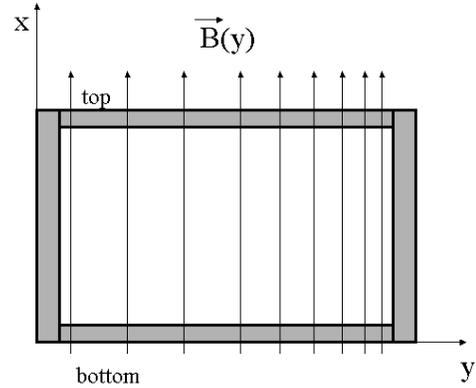, width=9cm}
\end{center}
\caption{Toy model of the magnetar core. Perfectly conducting incompressible fluid is sandwiched
between perfectly conducting top and bottom plates. Magnetic field---shown by thin arrows---is 
vertically directed and is a function of $y$ only. The z-axis is perpendicular to the page.}
\end{figure}
The magnetic field $\vec{B}(y)$ is  vertically directed and is a function of $y$ only,
threading both top and bottom plates.
Gravity is zero, and magnetic pressure gradient is trivially compensated
by the fluid pressure gradient. Consider now a $z$-independent Lagrangian
displacement $\zeta(x,y)$ in the $z$ direction. The equation of motion is
\begin{equation}
{\partial^2\zeta\over \partial t^2}=c^2(y){\partial^2\zeta\over \partial x^2}-
\gamma{\partial\zeta\over \partial t},
\label{em}
\end{equation}
where $c(y)=\sqrt{T(y)/\rho(y)}$ and $\gamma$ is a small damping
constant added for generality;
here 
$T=B^2/(4\pi)$ is the magnetic tension (in a superconducting fluid
$T=BB_{\rm crit}/(4\pi)$) and we have allowed $y$-dependence of the 
fluid density
$\rho$.
Clearly, Eq.~(\ref{em}) separates in $x$ and $y$, with eigenfunctions
\begin{equation}
\zeta_{ny_0}(x,y)=\sin(n\pi x/l_x) \delta(y-y_0) \exp(i\omega_{ny_0}t),
\label{zetany0}
\end{equation}
with eigenfrequencies
\begin{equation}
\omega_{ny_0}=\pm\sqrt{(n\pi/l_x)^2T(y_0)/\rho(y_0)-\gamma^2/4}+i\gamma/2.
\label{omega}
\end{equation}
Here $l_x$ is the box height, $n $ is the integer number, and $y_0$ is the
$y$-coordinate where the eigenfunction is localized. {\bf The spectrum is
continuous!}

 It is notable that for a spherical shell
numerical results
of RRR indicate that the spectrum of toroidal modes is also continuous [and the eigenmodes
are singular, as in Eq.~(\ref{zetany0}) above]. The continuity of part of the 
spectrum is a topological
property, which should remain when external parameters (the shape of the box,
density profile, etc) change continuously. Since the basic  MHD physics in our
example, in the fluid shell of RRR, and in the magnetar core are
similar, we believe  that the magnetar core 
supports a continuum of MHD modes.

We started this section by asking whether a global magnetar MHD-elastic  mode
can be responsible for an observed QPO. 
We can now see that such a mode of the whole magnetar will 
be coupled to a continuum
of the MHD modes in the core.
Experience from Quantum Mechanics tells us that a mode coupled to a continuum of
other modes decays if its frequency lies within the continuum (cf.~Fermi's 
Golden Rule). This is likely the situation here: the QPO frequencies are 
tens of Hertz, above the base Alfven-mode frequency (but obviously below that 
of infinite
number of higher-order Alfven modes). Let us 
illustrate explicitly the decay of a 
global mode by returning to the simple example in Figure 1. We now model the
presence of a crust by assuming that the top plate is allowed to move
in z-direction but is also a harmonic oscillator
(with restoring force provided by some external spring).
The equation of motion of the top plate is given by
\begin{equation}
{d^2Z\over dt^2}=-\omega_0^2Z-F/M,
\label{plate}
\end{equation}
where $Z$ is the top plate's displacement, $\omega_0$ is the proper frequency of the
oscillator, $F$ is the force due to magnetic stresses from the fluid below, and 
$M$ is the plate's mass. 
When the top plate participates in a global mode of frequency 
$\omega$, then $Z=Z_0\exp(i\omega t)$. It is straightforward, by using 
Equation (\ref{em}), to solve for the fluid motion:
\begin{equation}
\zeta(x,y)={\sin[k(y)x]\over \sin[k(y)l_x]}Z,
\label{zeta1}
\end{equation}
where 
\begin{equation}
k(y)=\sqrt{\omega^2-i\omega\gamma}/c(y)\simeq {\omega\over c(y)}
\left(1-i{\gamma\over 2\omega}\right).
\label{k}
\end{equation}
The back-reaction force $F$ can now be computed:
\begin{eqnarray}
F&=&-\int T(y)\left({\partial\zeta\over \partial x}\right)_{x=l_x}
dy dz\nonumber\\
&=&-l_z\zeta\int T(y)k(y) \cot[k(y)l_x] dy\label{F}
\end{eqnarray}
where $l_z$ is the z-dimension of the top plate.
Let $\gamma=0$. Then if $\omega$ real and is in resonance with at least one
of the continuum modes then $\cot{kl_x}$ will have a pole singularity
at some $y=y_0$
in the range of integration. 
The singularity is lifted if $\gamma$ is non-zero; in the limit of 
$\gamma\ll\omega$ the integration will produce an imaginary component of
the amplitude of $F$:
\begin{equation}
\Im[F\exp(-i\omega t)]=-l_z\zeta_0 {T(y_0)k(y_0)\pi \over |l_x k^{\prime}(y_0)|}.
\label{Imf}
\end{equation}
Comparing the 
above equation with the Eq.~(\ref{plate}), we see that
our assumption that $\omega$ was real was not self-consistent.
Within an order-of-magnitude  the imaginary part of $\omega$ is 
\begin{equation}
\Im(\omega)\sim {Tl_z l_y\over l_x M\omega}.
\label{imomega}
\end{equation}
Substituting the parameters relevant for a magnetar ($B=10^{15}$,
$l_z=l_y=l_x=10^6$cm, $M=3\times 10^{31}$g, $\omega\sim 300\hbox{rad/s}$), we get
\begin{equation}
\Im(\omega)\sim 10\hbox{s}^{-1}.
\label{Imomega1}
\end{equation}
Even allowing for several corrections of order 1,
we see that a global mode is expected to decay within a second. This is much shorter
than the timescale on which the QPO is observed, $\sim 100$s. 

One can ask the following question: is it possible that a global mode is not coupled to
the MHD continuum in the core? After all, along with the MHD continuum the core can
probably support discrete MHD modes as well (such modes are seen in RRR's numerical
work). However, while some global mode may be decoupled from the continuum,
the mode of our interest must involve the motion of the
crust, in order to produce the observed QPO. Unless magnetic field in the core
has high degree of symmetry (which seems unlikely, especially if one believes that
it was generated by turbulent dynamo), this crustal motion is bound to couple the QPO-generating
global mode to
the continuum of the core's MHD modes via the field-lines threading the crust. Thus the decay
of a global mode seems generic.

\section{Magnetar: a tuning fork or wet spaghetti?}
Observations argue for the former: QPO frequencies are stable over tens of seconds,
something strongly indicative of a mechanical mode. However, we have shown that the theory
argues for the latter. Strong MHD crust-core coupling destroys the stability of a purely
crustal mode, and the presence of a continuum of MHD modes in the core 
makes a global crust-core mode decay within a second.
So what is the origin of the magnetar QPOs? It can still be mechanical, but only if the magnetic 
field geometry is very different from what has been assumed---for example, if the  magnetic
field is largely confined to the crust or has an extremely incoherent
poloidal component. Alternatively, the QPOs may have a magnetospheric
origin; this was mentioned as a possibility in Barat et al.~(1983), but has remained unexplored.

\section{acknowledgements}
We thank Chris Thompson and Tony Piro for valuable discussions. We also thank the referee for several  insightful
comments and suggestions.

\end{document}